\magnification \magstep1
\raggedbottom
\openup 2\jot
\voffset6truemm
\pageno=1
\headline={\ifnum\pageno=1\hfill\else
\hfill {\it New results in heat-kernel asymptotics on 
manifolds with boundary} 
\hfill \fi}
\def\II{{\rm 1\!\hskip-1pt I}}
\def\cstok#1{\leavevmode\thinspace\hbox{\vrule\vtop{\vbox{\hrule\kern1pt
\hbox{\vphantom{\tt/}\thinspace{\tt#1}\thinspace}}
\kern1pt\hrule}\vrule}\thinspace}
\centerline {\bf NEW RESULTS IN HEAT-KERNEL ASYMPTOTICS}
\centerline {\bf ON MANIFOLDS WITH BOUNDARY}
\vskip 1cm
\leftline {Giampiero Esposito}
\vskip 1cm
\noindent
{\it Istituto Nazionale di Fisica Nucleare, Sezione di Napoli,
Mostra d'Oltremare Padiglione 20, 80125 Napoli, Italy}
\vskip 0.3cm
\noindent
{\it Universit\`a di Napoli Federico II, Dipartimento di Scienze Fisiche,
Complesso Universitario di Monte S. Angelo, Via Cintia, 
Edificio G, 80126 Napoli, Italy}
\vskip 1cm
\noindent
{\bf Abstract.} A review is presented of some recent progress in
spectral geometry on manifolds with boundary: local boundary-value
problems where the boundary operator includes the effect of 
tangential derivatives; application of conformal variations and 
other functorial methods to the evaluation of heat-kernel 
coefficients; conditions for strong ellipticity of the
boundary-value problem; fourth-order operators on manifolds with
boundary; non-local boundary conditions in Euclidean quantum
gravity. Many deep developments in physics and mathematics are
therefore in sight.
\vskip 5cm
\noindent
{\it Fourth Workshop on Quantum
Field Theory under the Influence of External Conditions},
Leipzig, September 1998 (DSF preprint 98/23).
\vskip 100cm
\leftline {\bf 1. Introduction}
\vskip 0.3cm
\noindent
The investigation of the heat equation for differential 
(or pseudo-differential) operators on manifolds with boundary
remains of high relevance in physics and mathematics. Mathematicians
are more interested in the general properties of spectral geometry
and spectral asymptotics, with the aim of interpreting the various
heat-kernel coefficients with the help of invariance theory [1--3] in a
neat, elegant and deep way, and of finding resolvent and heat kernel
also when the boundary-value problem is pseudo-differential [4].
Heat-equation proofs of the index theorem are also available,
by now [3,5]. Theoretical physicists, on the other
hand, remain more interested in the applications, 
e.g. boundary conditions for Euclidean
quantum gravity [6--13], one-loop semiclassical approximation [8,12], 
and the quantization programme for gauge theories on manifolds
with boundary [14--17]. 

In the first part of our presentation, we shall consider an
$m$-dimensional Riemannian manifold, say $(M,g)$, a vector bundle
$V$ over $M$, with a connection $\nabla$, and operators of
Laplace type:
$$
P \equiv -g^{ab}\nabla_{a}\nabla_{b}-E ,
\eqno (1.1)
$$
with $E$ an endomorphism of $V$. The corresponding heat kernel
is, by definition, a solution, for $t>0$, of the equation
$$
\left({\partial \over \partial t}+P \right)U(x,x';t)=0,
\eqno (1.2)
$$
which obeys the initial condition
$$
\lim_{t \to 0} \int_{M}U(x,x';t)\rho(x')
\sqrt{{\rm det} \; g} \; dx=\rho(x) ,
\eqno (1.3)
$$
and suitable (local or non-local) boundary conditions
$$
\Bigr[{\cal B}U(x,x';t)\Bigr]_{\partial M}=0 \; \; \; \; .
\eqno (1.4)
$$
In the case of local boundary conditions, we shall assume them
to be of the form [17]
$$
\pmatrix{\Pi & 0 \cr \Lambda & \II-\Pi \cr}
\pmatrix{[\varphi]_{\partial M} \cr 
[\nabla_{N} \varphi]_{\partial M} \cr}=0,
\eqno (1.5)
$$
where $\Pi$ is a self-adjoint projection operator, $\Lambda$ is a 
tangential differential operator on the boundary of $M$:
$$
\Lambda \equiv (\II-\Pi)\left[{1\over 2}\Bigr(\Gamma^{i}
{\widehat \nabla}_{i}+{\widehat \nabla}_{i}\Gamma^{i}
\Bigr)+S \right](\II-\Pi),
\eqno (1.6)
$$
and $\varphi$ are the fields, i.e. the smooth sections of the
bundle $V$. With the notation of Eq. (1.6), $\widehat \nabla$ is
the induced connection on $\partial M$, $\Gamma^{i}$ are 
endomorphism-valued vector fields on the boundary, and $S$ is an
endomorphism of the vector bundle over $\partial M$ which is a
copy of $[V]_{\partial M}$, with sections given by 
$[\varphi]_{\partial M}$. $\Gamma^{i}$ and $S$ are anti-self-adjoint
and self-adjoint, respectively, and are annihilated by $\Pi$ on
the left and on the right, i.e. $\Pi \Gamma^{i}=\Gamma^{i} \Pi
=\Pi S = S \Pi=0$. As is shown in [3,11,12,16,17], one arrives
at such boundary conditions whenever one tries to obtain gauge-
and BRST-invariant boundary conditions in quantum field theory.

We will be interested in the asymptotic expansion as
$t \rightarrow 0^{+}$ of the $L^{2}$-trace [20] 
$$
{\rm Tr}_{L^{2}}\Bigr(fe^{-tP}\Bigr) \equiv \int_{M}
f(x) {\rm Tr}_{V} U(x,x;t)\sqrt{{\rm det} \; g} \; dx,
\eqno (1.7)
$$
where $f \in C^{\infty}(M)$. This is also called the 
{\it integrated heat kernel}. Equation (1.7) means that one first
takes the fibre trace of the heat-kernel diagonal. Composition
with the smearing function $f$, and integration over $M$, yields
the functional trace (1.7). The results for the original
boundary-value problem are eventually recovered by setting $f=1$,
but it is crucial to keep the smearing function arbitrary
throughout the whole set of calculations. In other words, we
consider the {\it global}, integrated asymptotics, for which
$$
{\rm Tr}_{L^{2}}\Bigr(f e^{-tP} \Bigr) \sim 
(4\pi t)^{-{m\over 2}} \sum_{n=0}^{\infty}t^{{n\over 2}}
A_{{n\over 2}}(f,P,{\cal B}).
\eqno (1.8)
$$
The coefficients $A_{{n\over 2}}$ can be always expressed in
the form
$$
A_{{n\over 2}}(f,P,{\cal B})=C_{{n\over 2}}(f,P)
+B_{{n\over 2}}(f,P,{\cal B}).
\eqno (1.9)
$$
The coefficients $C_{{n\over 2}}$ and $B_{{n\over 2}}$ are the
interior part and the boundary part, respectively. The interior
part vanishes for all odd values of $n$, whereas the boundary part
only vanishes if $n=0$. The interior part is obtained by integrating
over $M$ a linear combination of {\it local invariants} of the
appropriate dimension, built universally and polynomially from the
metric, the Riemann curvature $R_{\; bcd}^{a}$ of $M$, the bundle
curvature, say $\Omega_{ab}$, the endomorphism $E$ in the operator
(1.1), and their covariant derivatives. By virtue of the Weyl theorem
on the invariants of the orthogonal group [1,2], which is 
$O(m)$ for $M$, these polynomials can be found by using only tensor
products and contraction of tensor arguments. Moreover, the boundary
part is obtained upon integration over $\partial M$ of another linear
combination of local invariants. In that case, however, the structure
group is $O(m-1)$, and the coefficients of linear combination are
{\it universal functions}, independent of $m$, unaffected by 
conformal rescalings of the metric $g$, and invariant in form (i.e.
they are functions of position on the boundary, whose form is
independent of the boundary being curved or totally geodesic).

Thus, the general form of the $A_{{n\over 2}}$ coefficient is a
well posed problem in invariance theory, where one has to take all
possible local invariants built from $f, R_{\; bcd}^{a},\Omega_{ab},
K_{ij},E,S,\Gamma^{i}$ and their covariant derivatives (hereafter,
$K_{ij}$ is the extrinsic-curvature tensor of the boundary), 
integrating eventually their linear combinations over $M$ and
$\partial M$, respectively. For example, in the boundary part 
$B_{{n\over 2}}(f,P,{\cal B})$, the local invariants integrated
over $\partial M$ are of dimension $n-1$ in tensors of the same
dimension of the second fundamental form of the boundary, for all
$n \geq 1$. The universal functions associated to all such 
invariants can be found by using functorial methods, e.g. conformal
variations, lemmas on product manifolds [1,2], jointly with the
consideration of some particular manifolds (see below).

Section 2 shows how to apply the conformal-variation method to 
heat-kernel asymptotics for the generalized boundary-value
problem resulting from (1.1), (1.5) and (1.6). Section 3 describes
the recent results on the issue of strong ellipticity for this class
of local boundary-value problems. Fourth-order operators on 
manifolds with boundary are studied in section 4, and some recent
ideas on integro-differential boundary conditions in Euclidean
quantum gravity are discussed in section 5. Section 6 is devoted
to some open problems, and a number of new heat-kernel coefficients
are evaluated in the appendix.
\vskip 10cm
\leftline {\bf 2. Conformal variations and heat-kernel coefficients}
\vskip 0.3cm
\noindent
The conformal-variation method [18--21] is based on the behaviour
of the heat kernel under conformal rescalings. The idea is to
perform a conformal deformation of the differential operator $P$ 
and of the boundary operator $\cal B$, with a deformation parameter,
say $\varepsilon$ (see below). This is used to determine a set of
recurrence relations (see (2.1)--(2.3)) which reduce the evaluation
of heat-kernel asymptotics to the solution of a system of algebraic
equations for a finite set of coefficients. More precisely, the
``conformal variations" of the metric and of the inward-pointing
normal are $g_{ab}(\varepsilon)=e^{2 \varepsilon f}g_{ab}(0),
N^{a}(\varepsilon)=e^{-\varepsilon f}N^{a}(0)$. The conformal
variation of the potential terms $E(\varepsilon), S(\varepsilon)$
and $\Gamma^{i}(\varepsilon)$ is chosen in such a way that the
operator $P$ and the boundary operator $\cal B$ transform
according to $P(\varepsilon)=e^{-2 \varepsilon f}P(0),
{\cal B}(\varepsilon)=e^{-\varepsilon f}{\cal B}(0)$. This implies,
in particular, that $\Gamma^{i}(\varepsilon)=e^{-\varepsilon f}
\Gamma^{i}(0)$. Three basic conformal-variation formulae hold [1-3]:  
$$
\left[{d\over d\varepsilon}A_{{n\over 2}}\Bigr(1,
e^{-2 \varepsilon f}P(0),e^{-\varepsilon f}{\cal B}(0)\Bigr)
\right]_{\varepsilon=0}=(m-n)A_{{n\over 2}}
\Bigr(f,P(0),{\cal B}(0)\Bigr),
\eqno (2.1)
$$
$$
\left[{d\over d\varepsilon}A_{{n\over 2}}
\Bigr(1,P(0)-\varepsilon H,e^{-\varepsilon f}{\cal B}(0)
\Bigr)\right]_{\varepsilon=0}=A_{{n\over 2}-1}
\Bigr(H,P(0),{\cal B}(0)\Bigr),
\eqno (2.2)
$$
$$
\left[{d\over d\varepsilon}A_{{n\over 2}}\Bigr(
e^{-2 \varepsilon f}H, e^{-2 \varepsilon f} P(0),
e^{-\varepsilon f}{\cal B}(0)\Bigr)\right]_{\varepsilon=0}=0,
\eqno (2.3)
$$
where $H$ is another function $\in C^{\infty}(M)$, different from
$f$. Our analysis, relying on Refs. [20,21], is restricted to the 
case when the following conditions hold:
$$
\Bigr[\Gamma^{i},\Gamma^{j}\Bigr]=0,
\eqno (2.4)
$$
$$
\Bigr[\Gamma^{2},S \Bigr]=0,
\eqno (2.5)
$$
$$
{\widehat \nabla}_{i}\Gamma^{j}=0.
\eqno (2.6)
$$
In other words, the matrices $\Gamma^{i}$ commute with each other,
the matrix $\Gamma^{2} \equiv \Gamma_{i} \Gamma^{i}$ commutes
with $S$, and the matrices $\Gamma^{i}$ are taken to be covariantly
constant with respect to the induced connection, 
$\widehat \nabla$, on the boundary. These assumptions are indeed
quite restrictive, but, for the time being, not even the $A_{1}$
coefficient is known if (2.4)--(2.6) fail to hold, because formidable
technical difficulties are found to arise [17,20,21]. Thus, Eqs.
(2.4)--(2.6) do not express a compulsory step, but represent the
current limits of what one is able to do.

Although the Eqs. (2.4)--(2.6) impose severe restrictions, a
non-trivial invariance theory is found to contribute many new local
invariants to the heat-kernel asymptotics. Explicitly, one finds
the following structure of the first few interior terms [18,20,21]:
$$
B_{{1\over 2}}(f,P,{\cal B})=\sqrt{4\pi} 
\int_{\partial M}{\rm Tr}(\gamma f),
\eqno (2.7)
$$
$$
B_{1}(f,P,{\cal B})={1\over 6} \int_{\partial M}{\rm Tr}
\biggr[f(b_{0}K+b_{2}S)+b_{1}f_{;N}
+f \Bigr(\sigma_{1}K_{ij}\Gamma^{i}\Gamma^{j}\Bigr)\biggr],
\eqno (2.8)
$$
$$ \eqalignno{
\; & B_{{3\over 2}}(f,P,{\cal B})={\sqrt{4\pi}\over 384}
\int_{\partial M}{\rm Tr} \left \{ f \biggr[\Bigr(c_{0}E
+c_{1}R+c_{2}R_{\; \; \; \; iN}^{iN}+c_{3}K^{2} \right . \cr
&\left . +c_{4}K_{ij}K^{ij}+c_{7}SK+c_{8}S^{2}\biggr]
+f_{;N}(c_{5}K+c_{9}S)+c_{6}f_{;NN} \right \} \cr
&+{\sqrt{4\pi}\over 384} \int_{\partial M}{\rm Tr}
\left \{ f \biggr[\sigma_{2}(K_{ij}\Gamma^{i}\Gamma^{j})^{2}
+\sigma_{3}K_{ij}\Gamma^{i}\Gamma^{j}K
+\sigma_{4}K_{il}K_{\; j}^{l} \Gamma^{i} \Gamma^{j} \right . \cr
&+\lambda_{1}K_{ij}\Gamma^{i}\Gamma^{j}S
+\mu_{1}R_{i \; \; \; jN}^{\; \; N}\Gamma^{i}\Gamma^{j}
+\mu_{2}R_{\; ilj}^{l}\Gamma^{i}\Gamma^{j} \cr
& \left . +{\tilde b}_{1}\Omega_{iN}\Gamma^{i} \biggr]
+\beta_{1} f_{;N}K_{ij}\Gamma^{i}\Gamma^{j} \right \} .
&(2.9)\cr}
$$
For any subsequent interior term, the number of new local invariants
resulting from the occurrence of $\Gamma^{i}$ in the boundary
operator is higher and higher (but finite). For example, in the 
integrand for the coefficient $B_{2}(f,P,{\cal B})$, 33 new local
invariants (involving $\Gamma^{i}$) multiply $f$, 7 local invariants 
multiply $f_{;N}$, and 1 local invariant,  
$K_{ij}\Gamma^{i}\Gamma^{j}$, multiplies $f_{;NN}$ [20,21]. 
All universal functions occurring in (2.7)--(2.9) have been 
evaluated, thanks to the work in Refs. [15,17,18,20,21]. They are 
all expressed in terms of the functions
$$
\sqrt{1+\Gamma^{2}}, \; \; \sqrt{-\Gamma^{2}}, \; \; 
{\rm Arctanh}(\sqrt{-\Gamma^{2}}).
$$
For example, one finds, with the help of Eq. (2.1), and considering
a particular case (i.e. a flat background with a totally geodesic
boundary) which provides further information [18,20]:
$$
\sigma_{1}={6\over \Gamma^{2}}\left({1\over \sqrt{-\Gamma^{2}}}
{\rm Arctanh}(\sqrt{-\Gamma^{2}})
-{1\over (1+\Gamma^{2})} \right).
\eqno (2.10)
$$
A remarkable piece of work, in Ref. [21], has evaluated all universal
functions in (2.9), by using the formulae (2.1)--(2.3), jointly
with a lemma on product manifolds [1,2], and the consideration of
particular manifolds, i.e. the bounded generalized cone, and the 
manifold with $B^{2} \times T^{m-2}$ topology ($T^{m-2}$ being the
equilateral $(m-2)$-dimensional torus). The consideration 
of particular cases is always useful and, indeed, strictly necessary
(so far), by virtue of the universal nature of the functions of
$\Gamma^{2}$ one is looking for (see comments after (1.9)).
\vskip 0.3cm
\leftline {\bf 3. Strong ellipticity}
\vskip 0.3cm
\noindent
If one studies compact Riemannian manifolds without boundary, it
is enough to make sure that the leading symbol of the differential
operator under consideration is elliptic. In the presence of
boundaries, however, one has also to check that the strong
ellipticity condition holds. At a technical level, this requires 
that a unique solution should exist of the eigenvalue equation for
the leading symbol $\sigma_{L}(P)$, subject to a decay condition
at infinity and to suitable boundary conditions. To obtain a precise
formulation, one has to consider local coordinates on the boundary
$\partial M$, say ${\hat x}^{k}$ ($k=1,...,m-1$), the normal 
geodesic distance to $\partial M$, say $r$, cotangent vectors on the
boundary, say $\zeta_{j} \in T^{*}({\partial M})$ ($j=1,...,m-1$), a
real parameter, say $\omega$, the {\it graded leading symbol} of the 
boundary operator in Eq. (1.5) (cf. the first matrix therein):
$$
\sigma_{g}(B_{P}) \equiv \pmatrix{\Pi & 0 \cr 
i \Gamma^{j}\zeta_{j} & \II -\Pi \cr},
\eqno (3.1)
$$
and an arbitrary pair of boundary data, i.e.
$$
\psi' \equiv \pmatrix{\psi_{0}' \cr \psi_{1}' \cr},
\eqno (3.2)
$$
where such a $\psi'$ is, strictly, a smooth section of an
``auxiliary" vector bundle over $\partial M$, endowed with a
decomposition into sub-bundles, with half the dimension of the
bundle of boundary data.

By definition, strong ellipticity holds with respect to the
cone ${\cal C}-{\cal R}_{+}$ if a unique solution exists, say
$\varphi$, of the equation [2,17,22]
$$
\Bigr[\sigma_{L}(P;{\hat x}^{k},r=0,\zeta_{j},-i\partial_{r})
-\lambda \II \Bigr]\varphi(r)=0,
\eqno (3.3)
$$
subject to the asymptotic condition
$$
\lim_{r \to \infty} \varphi(r)=0,
\eqno (3.4)
$$
and to the boundary condition (hereafter, ${\hat x} \equiv
\left \{ {\hat x}^{k} \right \} ,
\zeta \equiv \left \{ \zeta_{j} \right \}$)
$$
\sigma_{g}(B_{P})({\hat x},\zeta)
\pmatrix{[\varphi]_{\partial M} \cr
[\nabla_{N}\varphi]_{\partial M} \cr}
=\pmatrix{\psi_{0}' \cr \psi_{1}' \cr},
\eqno (3.5)
$$
$\forall \zeta \in T^{*}({\partial M}), \forall \lambda \in
{\cal C}-{\cal R}_{+}, \forall (\zeta, \lambda) 
\not = (0,0)$ and $\forall \psi '$. What is crucial, for the
physicists who are interested in the applications to quantum
field theory (as well as for mathematicians who might be more
interested in heat-kernel theory), is that the lack of strong
ellipticity leads to a fibre trace of the heat-kernel diagonal
which acquires a non-integrable part near the boundary. It is
hence impossible to make sense of the integrated heat kernel
(cf. (1.7))
$$
{\rm Tr}_{L^{2}}(e^{-tP}) \equiv \int_{M}{\rm Tr}_{V}
U(x,x;t) \sqrt{{\rm det} \; g} \; dx ,
\eqno (3.6)
$$
with the corresponding global asymptotics as 
$t \rightarrow 0^{+}$ (unless one studies a smeared form along
the lines of (1.7)).

This is why, in Ref. [17], a systematic investigation of strong
ellipticity for local boundary-value problems involving operators
of Laplace and Dirac type has been carried out. For the former 
class of differential operators, which are our main source of
concern so far, the work in Refs. [17,22] may be summarized
as follows.
\vskip 0.3cm
\noindent
(i) Let $P$ be an operator of Laplace type, with boundary operator
as in Eq. (1.5):
$$
B_{P} \equiv \pmatrix{\Pi & 0 \cr \Lambda & \II-\Pi \cr}.
$$
The boundary-value problem $(P,B_{P})$ is strongly elliptic with
respect to ${\cal C}-{\cal R}_{+}$ if and only if, for all
$\zeta \not = 0$, the matrix $| \zeta | \II-i \Gamma^{j}\zeta_{j}$ is 
positive-definite, where $|\zeta| \equiv +\sqrt{\zeta_{j}\zeta^{j}}$.
\vskip 0.3cm
\noindent
(ii) The boundary-value problem for Euclidean Yang--Mills theory
at one-loop level, with the gauge-invariant boundary conditions
($N$ being the normal to the boundary)
$$
\Bigr[\left(\delta_{a}^{\; b}-N_{a}N^{b}\right)
\varphi_{b}\Bigr]_{\partial M}=0 ,
\eqno (3.7)
$$
$$
\Bigr[\nabla^{a}\varphi_{a}\Bigr]_{\partial M}=0,
\eqno (3.8)
$$
is strongly elliptic with respect to ${\cal C}-{\cal R}_{+}$.
\vskip 0.3cm
\noindent
(iii) In Euclidean quantum gravity at one-loop level, the vector
bundle $V$ is the bundle of symmetric rank-two tensor fields
$\varphi_{ab}$ over $M$, with fibre metric
$$
E^{ab \; cd} \equiv g^{a(c} \; g^{d)b}
+\alpha g^{ab}g^{cd},
\eqno (3.9)
$$
where $\alpha$ is a real parameter different from $-{1\over m}$
(for $\alpha=-{1\over m}$, no inverse of $E$ can be defined).
On considering the tensor $q_{ab} \equiv g_{ab}-N_{a}N_{b}$,
the projection operator which is self-adjoint with respect to the
bundle metric (3.9) reads
$$
\Pi_{ab}^{\; \; \; cd} \equiv q_{(a}^{\; c} \; q_{b)}^{\; d}
-{\alpha \over (\alpha+1)}N_{a}N_{b}q^{cd}.
\eqno (3.10)
$$
The boundary conditions invariant under infinitesimal 
diffeomorphisms on metric perturbations can be therefore expressed
in the form [17]
$$
\Bigr[\Pi_{ab}^{\; \; \; cd} \; \varphi_{cd}\Bigr]_{\partial M}=0,
\eqno (3.11)
$$
$$
\Bigr[E^{ab \; cd}(\alpha) \; \nabla_{b} \varphi_{cd}
\Bigr]_{\partial M}=0.
\eqno (3.12)
$$
If $\alpha=-{1\over 2}$, the operator on metric perturbations is
then of Laplace type, say $P$ again. Regrettably, one then finds
that the boundary-value problem $(P,B_{P})$, with $B_{P}$ the
boundary operator of the type (1.5) and (1.6) which gives rise
to (3.11) and (3.12) with $\alpha=-{1\over 2}$, is not strongly
elliptic with respect to ${\cal C}-{\cal R}_{+}$. The work in
Ref. [17] has also evaluated the non-integrable contribution
to the fibre trace of the heat-kernel diagonal, which can be
expected on general ground if strong ellipticity is violated,
as we said after Eq. (3.5).

In other words, only for Euclidean quantum gravity based on the
Einstein action the strong ellipticity condition is violated, on
using gauge-invariant boundary conditions of the form (3.11)
and (3.12). On the other hand, as shown in Ref. [16], such
boundary conditions can be re-derived, independently, by requiring
invariance under BRST transformations. It therefore seems that a
BRST-invariant quantization of the gravitational field presents
serious inconsistencies on manifolds with boundary, unless one 
accepts the view according to which the BRST invariance of the
amplitudes does not force the boundary conditions to be, themselves,
BRST invariant [14].
\vskip 0.3cm
\leftline {\bf 4. Fourth-order operators on manifolds
with boundary}
\vskip 0.3cm
\noindent
The current attempts to develop quantum field theories of 
fundamental interactions have led to the consideration of
fourth-order or even higher-order differential operators on
closed Riemannian manifolds [23--27], or on manifolds with
boundary [3,28,29]. The analysis of the transformation properties
under conformal rescalings of the background metric, say $g$,
leads, in particular, to the consideration of conformally
covariant operators, say $Q$, which transform according
to the law
$$
Q_{\omega}=e^{-(m+4)\omega /2}Q(\omega=0)
e^{(m-4)\omega /2} ,
\eqno (4.1)
$$
if $g$ rescales as $g_{\omega}=e^{2\omega}g$, $m$ being
the dimension of the Riemannian manifold which is studied. One
of the physical motivations for this analysis lies in the
possibility to use the Green functions of such operators to
build the effective action in curved space-times [25].

Another enlightening example is provided 
by the ghost sector of Euclidean Maxwell theory in
vacuum in four dimensions. The corresponding field equations are
well known to be invariant under conformal rescalings of $g$. 
On the other hand, the supplementary
(or gauge) conditions usually considered in the literature are
not invariant under conformal rescalings of $g$. Even just in
flat Euclidean four-space, conformal invariance of the supplementary
condition is only achieved on making the 
Eastwood--Singer choice [30]:
$$
\nabla_{b}\nabla^{b}\nabla^{c}A_{c}=0,
\eqno (4.2)
$$
where $A_{c}$ is the electromagnetic potential (a connection
one-form in geometric language). The preservation of Eq. (4.2) under
gauge transformations of $A_{c}$:
$$
{ }^{f}A_{c} \equiv A_{c}+\nabla_{c}f ,
\eqno (4.3)
$$
is achieved provided that $f$ obeys the fourth-order equation
$$
\cstok{\ }^{2}f=0 ,
\eqno (4.4)
$$
where $\cstok{\ }^{2}$ is the box operator composed with itself:
$$
\cstok{\ }^{2} \equiv \nabla_{a}\nabla^{a}\nabla_{b}\nabla^{b}.
$$
In the corresponding quantum theory via path integrals, one thus
deals with two independent ghost fields (frequently referred to
as the ghost and the anti-ghost), both ruled by $\cstok{\ }^{2}$,
which is a fourth-order elliptic operator, and subject to the
following boundary conditions: 
$$
[\varepsilon]_{\partial M}=0,
\eqno (4.5)
$$
$$
\Bigr[\nabla_{N}\varepsilon \Bigr]_{\partial M}=0 .
\eqno (4.6)
$$
Remarkably, since one now deals with a fourth-order elliptic
operator, it is insufficient to impose just Dirichlet or
Neumann (or Robin) boundary conditions. One needs instead both
(4.5) and (4.6), which are obtained from the following 
requirements: 
\vskip 0.3cm
\noindent
(i) Gauge invariance of the boundary conditions on 
$A_{b}$ [3,28]. 
\vskip 0.3cm
\noindent
(ii) Conformal invariance of the whole set of boundary 
conditions. 
\vskip 0.3cm
\noindent
(iii) Self-adjointness of the $\cstok{\ }^{2}$ operator 
(see below).
\vskip 0.3cm
\noindent
Although it remains extremely difficult to build a consistent 
quantization scheme via path-integral formalism for the full
Maxwell field in the Eastwood--Singer gauge (the gauge-field
operator on $A_{b}$ perturbations being, then, of sixth order
[3,28]), the investigation of the ghost sector remains of considerable
interest in this case. There is in fact, on the one hand, the need
to understand how to quantize a gauge theory in a way which
preserves conformal invariance at all stages (as we just said),
and on the other hand the attempt to extend the recent work on 
conformally covariant operators [23--27] to the more realistic case 
of manifolds with boundary.

For simplicity, we consider the squared Laplace operator acting on
scalar fields on a flat Euclidean background, in the case when
curvature effects result from the boundary only. Moreover,
motivated by quantum cosmology and Euclidean quantum gravity,
the boundary is assumed to be a three-sphere of radius $a$, say,
or a pair of concentric three-spheres [8,12]. The former case, in
particular, may be viewed as the limiting case when the wave
function of the universe is studied at small three-geometries
(i.e. as $a \rightarrow 0$), as shown in Ref. [31].

In our problem it is hence possible to expand
the scalar field on a family of three-spheres centred on the origin,
according to the familiar relation [32]
$$
\varepsilon(x,\tau)=\sum_{n=1}^{\infty}\varepsilon_{n}(\tau)
Q^{(n)}(x) ,
\eqno (4.7)
$$
where $\tau \in [0,a]$, $Q^{(n)}$ are the scalar harmonics on a
unit three-sphere, $S^{3}$, and $x$ are local coordinates on
$S^{3}$. Thus, one is eventually led to study a one-dimensional
differential operator of fourth order, and this makes it clear 
why all the essential information is obtained by the analysis
of the operator $B \equiv {d^{4}\over dx^{4}}$ on a closed interval
of the real line, say $[0,1]$. The operator $B$ is required to act
on functions which are at least of class $C^{4}$ (see (4.18)),
and the following definition of scalar
product (anti-linear in the first argument) is considered:
$$
(u,v) \equiv \int_{0}^{1}u^{*}(x)v(x)dx .
\eqno (4.8)
$$
We now want to study under which conditions the operator $B$ is
self-adjoint, which means that it should be symmetric, and its
domain should coincide with the domain of the adjoint, say
$B^{\dagger}$. For this purpose, we first study the relation
between the scalar products $(Bu,v)$ and $(u,Bv)$. We have then
to integrate repeatedly by parts, using the Leibniz rule
to express
$$
{d\over dx}\left({d^{3}u^{*}\over dx^{3}}v \right), \;
{d\over dx}\left({d^{2}u^{*}\over dx^{2}}{dv\over dx}
\right), \;
{d\over dx}\left({du^{*}\over dx}{d^{2}v\over dx^{2}}
\right), \;
{d\over dx}\left(u^{*}{d^{3}v\over dx^{3}}\right) .
$$
This leads to
$$ 
(Bu,v)=\left[{d^{3}u^{*}\over dx^{3}}v \right]_{0}^{1}
-\left[{d^{2}u^{*}\over dx^{2}}{dv\over dx}\right]_{0}^{1}
+\left[{du^{*}\over dx}{d^{2}v\over dx^{2}}\right]_{0}^{1} 
-\left[u^{*}{d^{3}v \over dx^{3}}\right]_{0}^{1}
+(u,Bv) .
\eqno (4.9)
$$
Bearing in mind that the adjoint, $B^{\dagger}$, of
${d^{4}\over dx^{4}}$ is again the operator 
${d^{4}\over dx^{4}}$, it is thus clear that the condition
$(Bu,v)=(u,B^{\dagger}v)$ is fulfilled provided that {\it both}
$u \in D(B)$ {\it and} $v \in D(B^{\dagger})$ obey one of
the following four sets of boundary conditions:
\vskip 0.3cm
\noindent
(i) First option:
$$
u(0)=u(1)=0 \; \; \; \; u'(0)=u'(1)=0,
\eqno (4.10)
$$
$$
v(0)=v(1)=0 \; \; \; \; v'(0)=v'(1)=0 .
\eqno (4.11)
$$
\vskip 0.3cm
\noindent
(ii) Second option:
$$
u(0)=u(1)=0 \; \; \; \; u''(0)=u''(1)=0,
\eqno (4.12)
$$
$$
v(0)=v(1)=0 \; \; \; \; v''(0)=v''(1)=0 .
\eqno (4.13)
$$
\vskip 0.3cm
\noindent
(iii) Third option: 
$$
u'(0)=u'(1)=0 \; \; \; \; u'''(0)=u'''(1)=0,
\eqno (4.14)
$$
$$
v'(0)=v'(1)=0 \; \; \; \; v'''(0)=v'''(1)=0.
\eqno (4.15)
$$
\vskip 0.3cm
\noindent
(iv) Fourth option:
$$
u''(0)=u''(1)=0 \; \; \; \; u'''(0)=u'''(1)=0,
\eqno (4.16)
$$
$$
v''(0)=v''(1)=0 \; \; \; \; v'''(0)=v'''(1)=0 .
\eqno (4.17)
$$
In other words, if the conditions (4.10) and (4.11), or (4.12)
and (4.13), or (4.14) and (4.15),
or (4.16) and (4.17) are satisfied, the domains of
$B$ and of its adjoint turn out to coincide [29]: 
$$ \eqalignno{
D(B)&=D(B^{\dagger}) \equiv
\left \{u: u \in AC^{4}[0,1], \;
(4.10) \; {\rm or} \; (4.12) \right . \cr
&\left . \; {\rm or} \; (4.14) \;
{\rm or} \; (4.16) \; {\rm holds} \right \}.
&(4.18)\cr}
$$
With our notation, $AC^{4}[0,1]$ is the set of functions in
$L^{2}[0,1]$ whose weak derivatives up to third order are
absolutely continuous in [0,1], which ensures that the weak
derivatives, up to fourth order, are Lebesgue summable 
in [0,1], and that all $u$ in the domain are of class $C^{4}$
on [0,1]. Of course, symmetry of $B$ is also obtained with the 
boundary conditions just described.

In other words, four sets of boundary conditions, (i) or (ii) 
or (iii) or (iv), can be chosen
to ensure self-adjointness of the operator ${d^{4}\over dx^{4}}$.
Hereafter, we first consider the option (i), since
it agrees with boundary conditions motivated by
the request of gauge invariance and conformal invariance, if the
scalar field is viewed as one of the two ghost fields of 
Euclidean Maxwell theory in the Eastwood-Singer gauge. We also
stress again that nothing is lost on studying just the 
``prototype" operator ${d^{4}\over dx^{4}}$. The one-dimensional
fourth-order operator may take a more complicated form in some
set of local coordinates (see (4.19)), but is always
reducible to the operator ${d^{4}\over dx^{4}}$ on the real
line (more precisely, a closed interval of $\Re$ in our problems).

The definition and evaluation of functional determinants remains
a topic of crucial importance in quantum field theory. Here the
task is even more interesting, because we are studying a
fourth-order elliptic operator on a manifold with boundary. As
shown in Refs. [3,28], the resulting eigenvalue equation for the 
modes occurring in the expansion (4.7) turns out to be, on the
Euclidean four-ball,
$$ 
\biggr[{d^{4}\over d\tau^{4}}+{6\over \tau}{d^{3}\over d\tau^{3}}
-{(2n^{2}-5)\over \tau^{2}}{d^{2}\over d\tau^{2}} 
-{(2n^{2}+1)\over \tau^{3}}{d\over d\tau}
+{(n^{2}-1)^{2}\over \tau^{4}}\biggr]\varepsilon_{n}
=\lambda_{n}\varepsilon_{n}.
\eqno (4.19)
$$
Thus, on setting $M \equiv \lambda_{n}^{1/4}$, the solution of
Eq. (4.19) is expressed by a linear combination of Bessel 
functions and modified Bessel functions [29], i.e.
$$ 
\varepsilon_{n}(\tau)=A_{1,n}{I_{n}(M\tau)\over \tau}
+A_{2,n}{K_{n}(M\tau)\over \tau} 
+A_{3,n}{J_{n}(M\tau)\over \tau}
+A_{4,n}{N_{n}(M\tau)\over \tau}.
\eqno (4.20)
$$
Since the Euclidean four-ball consists of a portion of flat
Euclidean four-space bounded by a three-sphere, the coefficients
$A_{2,n}$ and $A_{4,n}$ have to vanish $\forall n \geq 1$,
to ensure regularity of $\varepsilon_{n}$ 
at the origin. One is thus left with
scalar modes of the form
$$
\varepsilon_{n}(\tau)=A_{1,n}{I_{n}(M\tau)\over \tau}
+A_{3,n}{J_{n}(M\tau)\over \tau}.
\eqno (4.21)
$$
These massless modes are subject to the boundary conditions
(see (4.5), (4.6) and (4.10), (4.11))
$$
[\varepsilon_{n}]_{\partial M}=0,
\eqno (4.22)
$$
$$
[d \varepsilon_{n}/d\tau ]_{\partial M}=0.
\eqno (4.23)
$$
The Eqs. (4.21)--(4.23) lead to the eigenvalue
condition (denoting by $a$ the radius of the three-sphere)
$$
{\rm det} \pmatrix{I_{n}(Ma)& J_{n}(Ma) \cr
-I_{n}(Ma)+Ma I_{n}'(Ma) & 
-J_{n}(Ma)+Ma J_{n}'(Ma) \cr}=0 ,
\eqno (4.24)
$$
which guarantees that non-trivial solutions exist for the
coefficients $A_{1,n}$ and $A_{3,n}$ in (4.21). As proved in
Ref. [29], Eq. (4.24) leads to the following $\zeta(0)$ value:
$$
\zeta(0)=-{1\over 120}.
\eqno (4.25)
$$
This is of some interest, because the one-loop analysis remains
crucial in the course of studying quantum theory as a theory of
small disturbances [33] of the underlying classical theory.

In the two-boundary problem one studies instead a portion of flat
Euclidean four-space bounded by two concentric three-spheres. This case
is very interesting because it is more directly related to the
familiar framework in quantum field theory, where one normally
assigns boundary data on two three-surfaces (it should be
stressed, however, that unlike scattering problems we are
considering a path-integral representation of amplitudes in
a finite region).
On denoting by $a$ and $b$, with $a > b$, the radii of the 
two concentric three-sphere boundaries, we can consider the
complete form (4.20) of our scalar modes, because no singularity
at the origin occurs in the two-boundary problem, and hence all
linearly independent integrals are regular, for all 
$\tau \in [b,a]$. We now impose the boundary conditions (4.5)
and (4.6), which lead to the eigenvalue condition
$$
{\rm det} \pmatrix{I_{n}(Mb) & K_{n}(Mb) & J_{n}(Mb)
& N_{n}(Mb) \cr
F_{I_{n}}(Mb) & F_{K_{n}}(Mb) & F_{J_{n}}(Mb)
& F_{N_{n}}(Mb) \cr
I_{n}(Ma) & K_{n}(Ma) & J_{n}(Ma) & N_{n}(Ma) \cr
F_{I_{n}}(Ma) & F_{K_{n}}(Ma) & F_{J_{n}}(Ma)
& F_{N_{n}}(Ma) \cr}=0 ,
\eqno (4.26)
$$
where, for $Z=I,K,J$ or $N$, we define [29]
$$
F_{Z_{n}}(Mx) \equiv -Z_{n}(Mx)+Mx Z_{n}'(Mx) .
\eqno (4.27)
$$
The work in Ref. [29] proves that Eq. (4.26) leads to
a vanishing $\zeta(0)$ value:
$$
\zeta(0)= 0 .
\eqno (4.28)
$$
The result (4.28) is found to hold for all boundary conditions
described in Eqs. (4.10)--(4.17).

To sum up, the original contribution of Ref. [29] is as follows.
\vskip 0.3cm
\noindent 
(i) The boundary conditions for which the squared Laplace 
operator is self-adjoint have been derived for the first time,
taking as prototype the operator ${d^{4}\over dx^{4}}$ on a
closed interval of the real line. Interestingly, four sets of
boundary conditions are then found to arise, and the option
described by (4.10) and (4.11) coincides, if the field in (4.7)
were a ghost field, with the boundary conditions obtained from
the request of gauge invariance of the boundary conditions on 
$A_{b}$, when the Eastwood--Singer supplementary condition
is imposed.
\vskip 0.3cm
\noindent
(ii) Given the fourth-order eigenvalue equation (4.19), the
contribution of the corresponding eigenmodes to the one-loop
divergence has been derived for the first time on the 
Euclidean four-ball (see (4.25)), or on the portion of flat 
Euclidean four-space bounded by two concentric three-spheres
(see (4.28)).
\vskip 0.3cm
\noindent
In our opinion, the result (i) is crucial because no complete
prescription for the quantization is obtained unless suitable
sets of boundary conditions are imposed, and Eqs. (4.10)--(4.17)
represent a non-trivial step in this direction.
The result (ii) is instead relevant for the analysis of
one-loop semiclassical effects in quantum field theory. In other
words, if one has to come to terms with higher order differential
operators in the quantization of gauge theories and gravitation,
it appears necessary to develop techniques for a systematic
investigation of one-loop ultraviolet divergences, as a
first step towards a thorough understanding of 
their perturbative properties.

Some outstanding problems are now in sight. First, it appears
interesting to extend our mode-by-mode analysis to curved
backgrounds with boundary. In this case, the fourth-order
conformally covariant differential operator is more complicated
than the squared Laplace operator, and involves also the
Ricci curvature and the scalar curvature of the background.
Second, one should use Weyl's theorem on the invariants of
the orthogonal group to understand the general structure of
heat-kernel asymptotics for fourth-order differential
operators on manifolds with boundary. A naturally occurring 
question within that framework is, to what extent functorial 
methods can then be used to compute all heat-kernel
coefficients for a given form of the differential operator
and of the boundary operator (cf. Refs. [20,21]). 
Third, the recently considered
effect of tangential derivatives in the boundary operator 
[15--18,20--22] might give rise to generalized boundary conditions
for conformally covariant operators. The appropriate 
mathematical theory is still lacking in the literature, but
would be of much help for the current attempts to understand
the formulation of quantum field theories on manifolds with
boundary. 
\vskip 0.3cm
\leftline {\bf 5. Non-local boundary conditions in Euclidean
quantum gravity}
\vskip 0.3cm
\noindent
The last decade of efforts on the problem of boundary conditions
in (one-loop) Euclidean quantum gravity has focused on a 
{\it local formulation}, by trying to satisfy the following 
requirements:
\vskip 0.3cm
\noindent
(i) Local nature of the boundary operators [6--9,11,12,16,17,20,22].
\vskip 0.3cm
\noindent
(ii) Operator on metric perturbations, say $P$, and ghost operator,
say $Q$, of Laplace type [17].
\vskip 0.3cm
\noindent
(iii) Symmetry, and, possibly, (essential) self-adjointness of the
differential operators $P$ and $Q$ [11,12].
\vskip 0.3cm
\noindent
(iv) Strong ellipticity of the boundary-value problems obtained
from the operators $P$ and $Q$, 
with local boundary operators $B_{1}$ and
$B_{2}$, respectively [17,22].
\vskip 0.3cm
\noindent
(v) Gauge- and BRST-invariance of the boundary conditions and/or
of the out-in (one-loop) amplitude [6,7,9,11,12,14,16,17,22].
\vskip 0.3cm
\noindent
At about the same time, in the applications to quantum field theory
and quantum gravity, non-local boundary conditions had been 
studied mainly for operators of Dirac type 
(see, however, Ref. [10]), 
relying on the early work by Atiyah, Patodi and Singer on spectral
asymmetry and Riemannian geometry [34]. What is non-local, within
that framework, is the separation of the spectrum of a first-order
elliptic operator (the Dirac operator on the boundary) into its
positive and negative parts. This leads, in turn, to an unambiguous
identification of positive- and negative-frequency modes of the
(massive or massless) Dirac field, and half of them are set to zero 
on the bounding surface [3,8,35].

On the other hand, non-local boundary conditions for operators of
Laplace type had already been studied quite intensively in the
literature, from at least two points of view:
\vskip 0.3cm
\noindent
(i) The rich mathematical theory of pseudo-differential
boundary-value problems, where both the differential operator $P$
and the boundary operator $B$ may be replaced by 
integro-differential operators [4].
\vskip 0.3cm
\noindent
(ii) Bose--Einstein condensation models, where integro-differential
boundary operators lead to the existence of bulk and
surface states [36].
\vskip 0.3cm
\noindent
For example, if $P$ is an operator of Laplace type, mathematicians
have derived many properties of the boundary-value problem [4]
$$
Pu=f \; \; \; \; {\rm in} \; \Omega ,
\eqno (5.1)
$$
$$
Tu=\varphi \; \; \; \; {\rm at} \; 
\Gamma \equiv \partial \Omega ,
\eqno (5.2)
$$
where the boundary operator $T$ can take the form
$$
Tu=\gamma_{0}u+T_{0}'u ,
\eqno (5.3)
$$
or, instead, 
$$
Tu=\gamma_{1}u+S_{0}\gamma_{0}u+T_{1}'u .
\eqno (5.4)
$$
With this notation, one has [4]
$$
\gamma_{j}u \equiv \Bigr[ (-i \partial_{n})^{j} u
\Bigr]_{\partial \Omega} \; \; \; \; , 
j=0,1,... 
\eqno (5.5)
$$
where $\partial_{n}$ is the inward-pointing normal derivative.
Moreover, $T_{0}'$ and $T_{1}'$ are integral operators going
from $\Omega$ to $\partial \Omega$, and the map $S_{0}$ acts
on functions on $\partial \Omega$. 

In the case of the gravitational field, 
inspired by Eqs. (5.1)--(5.5),
we consider a scheme where the differential operator on metric
perturbations remains of Laplace type (as well as the ghost
operator), whereas the boundary conditions are of integro-differential
nature. This means that the full boundary operator, say 
$B_{ab}^{\; \; \; cd}$, may be expressed as the sum of a local
operator, say ${\widetilde B}_{ab}^{\; \; \; cd}$, obtained from
projectors and first-order differential operators 
(see (1.5) and (1.6)), and an
integral operator going from the background four-manifold, $M$, 
to its boundary $\partial M$, so that the boundary conditions read
$$
\Bigr[B_{ab}^{\; \; \; cd}h_{cd}(x)\Bigr]_{\partial M}
=\Bigr[{\widetilde B}_{ab}^{\; \; \; cd}h_{cd}(x)\Bigr]_{\partial M}
+\left[\int_{\partial M}T_{ab}^{\; \; \; cd}(x,x')
h_{cd}(x')d\Sigma' \right]_{\partial M}.
\eqno (5.6a)
$$
This notation is a bit too general. We may decide, following
DeWitt [37], that unprimed lower-case indices refer to the point
$x$ and primed lower-case indices refer to the point $x'$.
This leads to
$$
\Bigr[B_{ab}^{\; \; \; cd}h_{cd}(x)\Bigr]_{\partial M}
=\Bigr[{\widetilde B}_{ab}^{\; \; \; cd}h_{cd}(x)\Bigr]_{\partial M}
+\left[\int_{\partial M}T_{ab}^{\; \; \; c'd'}h_{c'd'}
d\Sigma' \right]_{\partial M},
\eqno (5.6b)
$$
which is the form of the boundary conditions 
chosen hereafter [13].

Since we are concerned, for simplicity, with operators of
Laplace type in a flat four-dimensional background (all
curvature effects result then from the boundary only), it is
very important for us to understand the effect of 
integro-differential boundary conditions on such a class of
operators. For this purpose, following Ref. [4], we remark
that, after integration by parts, one finds the Green formula
(unlike Ref. [4], we define the Laplace operator with a
negative sign in front of all second derivatives), for 
$P=\bigtriangleup$, $u \in D(P)$, and $v$ in the domain
$D(P^{*})$ of the adjoint of $P$:
$$
(Pu,v)_{\Omega}=\Bigr(\bigtriangleup u,v \Bigr)_{\Omega}
=\Bigr(u, \bigtriangleup v \Bigr)_{\Omega}
+\Bigr({\cal U}\rho u, \rho v \Bigr)_{\Gamma} ,
\eqno (5.7)
$$
where the {\it Green matrix} reads, in our case [4]
$$
{\cal U}=i \pmatrix{0 & I \cr I & 0 \cr} ,
\eqno (5.8)
$$
whilst $\rho$ is the (Cauchy) boundary operator, whose
action reduces to
$$
\rho u=(\gamma_{0}u, \gamma_{1}u).
\eqno (5.9)
$$
The same property (5.9) holds for $v \in D(P^{*})$. Suppose now
that the boundary conditions are expressed in the 
integro-differential form (5.3): 
$$
\gamma_{0}u+T_{0}'u=0 \; \; {\rm at} \; \; \Gamma .
$$
The term $\Bigr({\cal U}\rho u, \rho v \Bigr)_{\Gamma}$ in Eq. (5.7), 
which is equal to
$$
\Bigr({\cal U}\rho u, \rho v \Bigr)_{\Gamma}
=i (\gamma_{1}u,\gamma_{0}v)_{\Gamma}
+i (\gamma_{0}u,\gamma_{1}v)_{\Gamma} , 
\eqno (5.10a)
$$
can be then re-expressed as 
$$
\Bigr({\cal U}\rho u, \rho v \Bigr)_{\Gamma}
=i (\gamma_{1}u,\gamma_{0}v)_{\Gamma}
+i(-T_{0}'u,\gamma_{1}v)_{\Gamma} ,
\eqno (5.10b)
$$
which implies that $P^{*}$, the (formal) adjoint of $P$, can
be obtained by adding to $\bigtriangleup$ a singular Green
operator, i.e.
$$
P^{*}v=\bigtriangleup v + i T_{0}'^{*}\gamma_{1}v , 
\eqno (5.11)
$$
supplemented by the local boundary condition
$$
\gamma_{0}v=0 \; \; {\rm at} \; \; \Gamma.
\eqno (5.12)
$$
By contrast, if the boundary conditions (5.4) are imposed: 
$$
\gamma_{1}u+S_{0}\gamma_{0}u+T_{1}'u=0 \; \; 
{\rm at} \; \; \Gamma,
$$
which modify the standard Neumann case, it is convenient to
re-express $\gamma_{1}u$, at the boundary, in the form
$$
\gamma_{1}u=-S_{0}\gamma_{0}u-T_{1}'u ,
\eqno (5.13)
$$
and insert Eq. (5.13) into Eq. (5.10{\it a}). This implies that the
adjoint of $P$ now reads
$$
P^{*}v=\bigtriangleup v + i T_{1}'^{*} \gamma_{0}v , 
\eqno (5.14)
$$
subject to the local boundary condition
$$
\gamma_{1}v=0 \; \; {\rm at} \; \; \Gamma .
\eqno (5.15)
$$
In other words, we are discovering a property which is known
to some mathematicians, but not so familiar to physicists:
if an elliptic differential operator (here taken to be
of Laplace type) is studied with
integro-differential boundary conditions, its adjoint is a
pseudo-differential operator, subject to local 
boundary conditions.

Self-adjointness problems are properly formulated by studying 
the so-called {\it realization} of the operator $P$ [4]. In
our case, this means adding to the Laplacian a singular Green
operator, and considering a {\it trace operator} which expresses
the integro-differential boundary conditions. More precisely,
a Dirichlet-type realization of $P=\bigtriangleup$ 
is the operator
$$
B_{D} \equiv \Bigr(\bigtriangleup +G_{D} \Bigr)_{T_{0}} ,
\eqno (5.16)
$$
where
$$
G_{D} \equiv K_{1}\gamma_{1}+G' ,
\eqno (5.17)
$$
$$
T_{0} \equiv \gamma_{0}+T_{0}' .
\eqno (5.18)
$$
The $K_{i}$ operators, for $i=0,1$, are 
completely determined by the requirement of self-adjointness.
In technical language, they are called Poisson operators [4].
Indeed, the domains of $B_{D}$ and its adjoint coincide if and
only if [4]
$$
K_{1}=i \; T_{0}'^{*} , 
\eqno (5.19)
$$
$$
G'=G'^{*} .
\eqno (5.20)
$$
Moreover, a Neumann-type realization of $P=\bigtriangleup$
is the operator
$$
B_{N} \equiv \Bigr( \bigtriangleup + G_{N} \Bigr)_{T_{1}} , 
\eqno (5.21)
$$
where
$$
G_{N} \equiv K_{0}\gamma_{0}+F' , 
\eqno (5.22)
$$
$$
T_{1} \equiv \gamma_{1}+S_{0}\gamma_{0}+T_{1}' .
\eqno (5.23)
$$
The domains of $B_{N}$ and its adjoint are then found to coincide
if and only if [4]
$$
K_{0}=i \; T_{1}'^{*} , 
\eqno (5.24)
$$
$$
S_{0}=-S_{0}^{*} ,
\eqno (5.25)
$$
$$
F'=F'^{*} .
\eqno (5.26)
$$

In the case of the gravitational field, our boundary operator
(5.6{\it b}) corresponds to the 
integro-differential trace operator (5.18).
The local boundary operator ${\widetilde B}_{ab}^{\; \; \; cd}$
is taken to be the one for which the following conditions
are imposed on metric perturbations on a three-sphere boundary
of radius $a$ [7]:
$$
\Bigr[h_{ij}\Bigr]_{\partial M}=0 , 
\eqno (5.27)
$$
$$
\Bigr[h_{0i}\Bigr]_{\partial M}=0 ,
\eqno (5.28)
$$
$$
\left[{\partial h_{00}\over \partial \tau}
+{6\over \tau}h_{00}
-{\partial \over \partial \tau}(g^{ij}h_{ij})
\right]_{\partial M}=0 ,
\eqno (5.29)
$$
where $\tau \in [0,a]$. Equations (5.27)--(5.29) express, to our
knowledge, the only set of local boundary conditions which are
of Dirichlet type on $h_{ij}$ and $h_{0i}$, and for which strong
ellipticity of the boundary-value problem is not violated 
(cf. section 3).
Since we only want to modify the Dirichlet sector of such 
boundary conditions [13], which is expressed by (5.27) and (5.28), 
we have to require that (see (5.6{\it b}))
$$
T_{00}^{\; \; \; c'd'}=0 \; \; \forall c',d' .
\eqno (5.30)
$$
Thus, we eventually consider the operator
$$
{\cal A} \equiv \pmatrix{ \bigtriangleup + G \cr
S \cr} ,
\eqno (5.31)
$$
where $S$ is of the type (5.18) in its $ij$ and $0i$
components, i.e.
$$
\Bigr[S_{ij}^{\; \; \; cd}h_{cd}(x)\Bigr]_{\partial M}
=\Bigr[h_{ij}(x)\Bigr]_{\partial M}
+\left[\int_{\partial M}T_{ij}^{\; \; \; c'd'}h_{c'd'}
d\Sigma'\right]_{\partial M} , 
\eqno (5.32)
$$
$$
\Bigr[S_{0i}^{\; \; \; cd}h_{cd}(x)\Bigr]_{\partial M}
=\Bigr[h_{0i}\Bigr]_{\partial M}
+\left[\int_{\partial M}T_{0i}^{\; \; \; c'd'}h_{c'd'}
d\Sigma'\right]_{\partial M} ,
\eqno (5.33)
$$
and of the type (5.29) (cf. Eq. (5.23)) in its normal component
$h_{00}$, i.e.
$$
\Bigr[S_{00}^{\; \; \; cd}h_{cd}(x)\Bigr]_{\partial M}
=\left[{\partial h_{00}\over \partial \tau}
+{6\over \tau}h_{00}-{\partial \over \partial \tau}
(g^{ij}h_{ij})\right]_{\partial M} .
\eqno (5.34)
$$
Moreover, $\bigtriangleup$ is the standard Laplacian on
metric perturbations in flat Euclidean four-space, and $G$
may be viewed as the direct sum of $G_{D}$ and $G_{N}$
(cf. example 1.6.16 in Ref. [4]), with
$$
G_{D}=K_{1}\gamma_{1}+G' ,
\eqno (5.35)
$$
$$
G_{N}=F' ,
\eqno (5.36)
$$
subject to the self-adjointness conditions
$$
\Bigr[(K_{1})_{ij}^{\; \; \; cd}h_{cd}\Bigr]_{\partial M}
=i \left[\int_{\partial M}T_{ij}^{\; \; \; c'd'}h_{c'd'}
d\Sigma' \right]_{\partial M}^{*} , 
\eqno (5.37)
$$
$$
\Bigr[(K_{1})_{0i}^{\; \; \; cd}h_{cd}\Bigr]_{\partial M}
=i \left[\int_{\partial M}T_{0i}^{\; \; \; c'd'}h_{c'd'}
d\Sigma' \right]_{\partial M}^{*} ,
\eqno (5.38)
$$
$$
\Bigr[(K_{1})_{00}^{\; \; \; cd}h_{cd}\Bigr]_{\partial M}=0 ,
\eqno (5.39)
$$
$$
{G'}_{ab}^{\; \; \; cd}
=\left({G'}_{ab}^{\; \; \; cd}\right)^{*} ,
\eqno (5.40)
$$
$$
{F'}_{ab}^{\; \; \; cd}=\left({F'}_{ab}^{\; \; \; cd}\right)^{*}.
\eqno (5.41)
$$
Note that, by virtue of Eq. (5.30), the counterpart of $T_{1}'$
vanishes in our problem, and hence no counterpart of Eq. (5.24) 
has to be imposed. Further to this, the condition (5.25) is satisfied
by virtue of the boundary condition (5.34).
 
The contribution of Ref. [13], summarized in the present
section, consists of the proposal that the
boundary conditions (5.6{\it b}) should be considered as a serious
candidate for non-local boundary conditions in Euclidean 
quantum gravity; moreover, the self-adjointness
conditions (5.37)--(5.41) have been derived, inspired by a careful
analysis of the results first derived in Ref. [4]. At least
four outstanding problems are now in sight:
\vskip 0.3cm
\noindent
(i) Can one build explicitly a class of bulk and surface states
in Euclidean quantum gravity with non-local boundary conditions,
inspired by the work in Ref. [36]? The idea is then to obtain 
mode-by-mode solutions of the eigenvalue equations for metric
perturbations, and insert them into Eq. (5.6{\it b}) for a given
form of $T_{ab}^{\; \; \; c'd'}$. One then looks for solutions
which decay rapidly away from the boundary (the surface states),
or remain non-negligible (the bulk states).
\vskip 0.3cm
\noindent
(ii) Can one study heat-kernel asymptotics with non-local
boundary conditions for the gravitational field?
\vskip 0.3cm
\noindent
(iii) Can one perform a path-integral quantization, if the
non-local choice (5.6{\it b}) is made for the boundary data?
What form of boundary conditions should be imposed 
on ghost fields?
\vskip 0.3cm
\noindent
(iv) Can one obtain a non-local formulation of Euclidean
quantum gravity, where both the differential operators and
the boundary operators are replaced by suitable classes of
pseudo-differential operators, taking full advantage of the
recent progress in the field of functional calculus for
pseudo-differential boundary-value problems [4]?
\vskip 0.3cm
\leftline {\bf 6. Concluding remarks}
\vskip 0.3cm
\noindent
Our presentation, although incomplete for obvious length limits,
has tried to open the window on what we regard as several promising
research lines for the years to come. They can be summarized by a
number of questions, as follows.
\vskip 0.3cm
\noindent
(i) When tangential derivatives occur in the boundary operator,
can one compute further heat-kernel coefficients for operators
of Laplace type in the Abelian (see (2.4) and (2.5)) 
and non-Abelian cases?
\vskip 0.3cm
\noindent
(ii) What about the general structure of heat-kernel asymptotics
for non-minimal operators [2,38,39] and conformally covariant 
operators [23,24] on manifolds with boundary?
\vskip 0.3cm
\noindent
(iii) Can one obtain a non-local formulation of Euclidean
quantum gravity with the help of functional calculus for
pseudo-differential boundary-value problems, along the lines
of Refs. [4] and [13]? Is there a corresponding heat-kernel
asymptotics? Can one build explicitly bulk and surface states 
in Euclidean quantum gravity?
\vskip 0.3cm
\noindent
(iv) What lesson should one learn from the recent proof that
there is lack of strong ellipticity in Euclidean quantum gravity,
upon choosing completely gauge-invariant boundary conditions
with gauge-field and ghost operators of Laplace type [17,22]?
Can one then define a smeared functional trace (cf. (1.7)) which
is not {\it ad hoc}? Is it correct to conclude that
BRST-invariant boundary conditions for the quantized gravitational
field turn out to be incompatible with the need for a well
defined elliptic theory?

The first two problems that we have selected might seem very 
technical, but their solution would lead to a substancial
advancement of knowledge in spectral geometry and in the general
theory of operators on manifolds. The remaining problems lie,
instead, at the very heart of any attempt to get a deeper 
understanding of Euclidean quantum gravity. The whole programme
might seem as either too ambitious or too abstract to the
pragmatic reader, but such a stand would be most unfortunate.
In our opinion, there are, by contrast, several good reasons 
for continuing advanced research along these lines, i.e.
\vskip 0.3cm
\noindent
(i) Quantum gravity via Euclidean path integrals makes it 
possible to deal with a framework where, quite naturally,
partition functions are defined. This is, in turn, crucial if
one wants to combine quantum theory, whose predictions are of
statistical nature [3], with general relativity.
\vskip 0.3cm
\noindent
(ii) Non-local properties are frequently met in the course of
studying modern quantum field theories via path-integral or 
canonical methods (cf. Ref. [40]). Thus, a non-local approach to 
Euclidean quantum gravity appears essential to complete the current
efforts.
\vskip 0.3cm
\noindent
(iii) Spectral geometry plays a crucial role if one tries to get
a thorough understanding of the one-loop semiclassical 
approximation [3,8,12]. In a space-time approach [41--44], such
an approximation provides the ``bridge" in between the classical
world and the (as yet unknown) full quantum theory (via path
integrals). At some stage, any alternative approach to quantum
gravity should be able to make contact with what one knows from a
perturbative evaluation of transition amplitudes [12].
\vskip 0.3cm
\noindent
(iv) So much has been learned from the heat-kernel approach to
index theory and to the theory of eigenvalues in Riemannian
geometry [2--4,12] that any new result may have a non-trivial
impact on Euclidean quantum gravity. Some readers might feel unhappy
with the idea of having to formulate the quantum theory in a
Euclidean framework, where no time-evolution exists, the 
differential operators are elliptic and the geometries are
Riemannian (rather than Lorentzian as in general relativity).
However, if one is interested in the most fundamental structures,
it is not bad news to realize that a framework exists where some
problems become well posed. From this point of view, 
one has to go ahead as long as possible with
Euclidean problems, before being sure that quantum theory needs
a Lorentzian framework. In other words, if a manifold picture 
retains an important role in quantum field theory, the problem
is open of whether one should regard the elliptic boundary-value
problems as the most fundamental tool. In the years to come,
hopefully, further progress in the topics discussed so far might
lead to a deeper vision.
\vskip 0.3cm
\leftline {\bf Appendix}
\vskip 0.3cm
\noindent
The consideration of special cases remains of considerable help
in the investigation of heat-kernel asymptotics, because the local
invariants in the integrand for heat-kernel coefficients are
multiplied by universal functions, whose form remains the same
for all smooth boundaries (see section 1). Thus, if one is able to
evaluate a subset of the general set of universal functions, one 
can use this result, jointly with the algebraic equations resulting
from (2.1)--(2.3) and from the application of lemmas on product 
manifolds [1,2], until one eventually 
gets enough algebraic equations
to evaluate all universal functions corresponding to a given
heat-kernel coefficient [21].

Thus, following Ref. [20], we consider the case when all curvatures
vanish: $R_{\; bcd}^{a}=0, \Omega_{ab}=0, K_{ij}=0$, whilst
$E,S$ and $\Gamma^{i}$ are covariantly constant, in that
$$
\nabla E=0, \; {\widehat \nabla}S=0, \;
{\widehat \nabla}_{i}\Gamma^{j}=0 .
\eqno ({\rm A}.1)
$$
The coefficient $A_{5/2}$ is a purely boundary term, and from
Ref. [20] one finds in our case
$$
A_{5/2}=\int_{\partial M}{\rm Tr}_V\left[\sum_{k=0}^{2}
\sum_{j=0}^{4-2k}{1\over k!} \rho_{j,4-2k}(\Gamma)E^{k}S^{j}
f^{(4-2k-j)}\right],
\eqno ({\rm A}.2)
$$
where $f^{(r)}$ is the normal derivative of $f$ of order $r$
(e.g., $f^{(2)}=f_{;NN}$), and $\rho_{n,k}$ are universal
functions generated by the following algorithm: 
$$
\alpha_{n,k}={1\over 2}\Gamma \left({{n+k}\over 2}+1 \right)
\int_{0}^{\infty}du {u^{k}\over [u^{2}+u+{1\over 4}
(1+\Gamma^{2})]^{{{n+k}\over 2}+1}} ,
\eqno ({\rm A}.3)
$$
$$
\rho_{0,k}={1\over k!}\left[{1\over 2}\Gamma
\left({{k+1}\over 2}\right)-\Gamma^{2}\alpha_{1,k}\right],
\eqno ({\rm A}.4)
$$
$$
\rho_{n,k}={1\over n!(k-n)!}\Bigr[2n \alpha_{n-1,k-n}
-\Gamma^{2} \alpha_{n+1,k-n}\Bigr].
\eqno ({\rm A}.5)
$$
All universal functions are well defined and analytic provided
that $1+\Gamma^{2} >0$. As remarked in Refs. [18,20] this reflects a
crucial property: when $\Gamma^{2}=-1$, the strong ellipticity 
of the boundary-value problem no longer holds. In our paper,
we always assume that $1+\Gamma^{2}$ is positive.

Now the explicit form of $A_{5/2}$ is, from Eq. (A.2), 
$$
\eqalignno{
A_{5/2}&=\int_{\partial M}{\rm Tr}_V \Bigr[f(\rho_{4,4}S^{4}
+\rho_{2,2}ES^{2}+{1\over 2}\rho_{0,0}E^{2})\cr
&+f_{;N}(\rho_{3,4}S^{3}+\rho_{1,2}ES)
+f_{;NN}(\rho_{2,4}S^{2}+\rho_{0,2}E)\cr
&+f_{;NNN} \; \rho_{1,4}S +f_{;NNNN} \; \rho_{0,4}\Bigr].
&({\rm A}.6)\cr}
$$
A careful application of the algorithm ({\rm A}.3)--({\rm A}.5) yields
therefore
$$
\rho_{0,0}=\sqrt{\pi}\left[-{1\over 2}+{1\over 
\sqrt{1+\Gamma^{2}}}\right],
\eqno ({\rm A}.7)
$$
$$
\rho_{2,2}={\sqrt{\pi}\over (1+\Gamma^{2})^{3/2}},
\eqno ({\rm A}.8)
$$
$$
\rho_{4,4}={1\over 2}{\sqrt{\pi}\over (1+\Gamma^{2})^{5/2}},
\eqno ({\rm A}.9)
$$
$$
\rho_{1,2}={\sqrt{\pi}\over \Gamma^{2}}
\left[1-{1\over \sqrt{1+\Gamma^{2}}}\right],
\eqno ({\rm A}.10)
$$
$$
\rho_{3,4}={\sqrt{\pi}\over 3 \Gamma^{4}}\left[2
-{(2+3 \Gamma^{2})\over (1+\Gamma^{2})^{3/2}}\right],
\eqno ({\rm A}.11)
$$
$$
\rho_{0,2}=\sqrt{\pi}\left[-{1\over 8}
+{1\over 2 \Gamma^{2}}\Bigr(\sqrt{1+\Gamma^{2}}
-1 \Bigr)\right],
\eqno ({\rm A}.12)
$$
$$
\rho_{2,4}=\sqrt{\pi}\left[{1\over \Gamma^{4}}
\Bigr(\sqrt{1+\Gamma^{2}}-1 \Bigr)
-{1\over 2}{1\over \Gamma^{2}\sqrt{1+\Gamma^{2}}}
\right],
\eqno ({\rm A}.13)
$$
$$
\rho_{1,4}=\sqrt{\pi}\left[{1\over 4}{1\over \Gamma^{2}}
+{1\over 2}{1\over \Gamma^{4}} \Bigr(1-\sqrt{1+\Gamma^{2}}
\Bigr)\right],
\eqno ({\rm A}.14)
$$
$$
\rho_{0,4}={\sqrt{\pi}\over 64}\left[-1-{8\over \Gamma^{2}}
+{16\over 3}{1\over \Gamma^{4}}\Bigr((1+\Gamma^{2})^{3/2}
-1 \Bigr)\right].
\eqno ({\rm A}.15)
$$
Note that all these new universal functions have a smooth
limit at $\Gamma=0$, where they coincide with the Robin
coefficients: ${\sqrt{\pi}\over 2},\sqrt{\pi},
{\sqrt{\pi}\over 2},{\sqrt{\pi}\over 2},
{\sqrt{\pi}\over 4},{\sqrt{\pi}\over 8},
{\sqrt{\pi}\over 8},-{\sqrt{\pi}\over 8},
{\sqrt{\pi}\over 64}$, respectively. 

These new universal
functions can be computed by hand after noticing that, by
virtue of (A.3)--(A.5), one deals repeatedly with the integral
$$
R(a,b,\Gamma) \equiv \int_{0}^{\infty}du 
{u^{a}\over [u^{2}+u+{1\over 4}(1+\Gamma^{2})]^{b}}.
\eqno ({\rm A}.16)
$$
For example, one finds
$$
R(4,7/2,\Gamma)=R(2,5/2,\Gamma)
-R(3,7/2,\Gamma)-{1\over 4}(1+\Gamma^{2})
R(2,7/2,\Gamma),
\eqno ({\rm A}.17)
$$
and this leads to the result (A.15), because
$$
R(2,5/2,\Gamma)=-{8\over 3}
{\sqrt{1+\Gamma^{2}}\over \Gamma^{4}}+{8\over 3}
{1\over \Gamma^{4}}+{4\over 3}{1\over \Gamma^{2}},
\eqno ({\rm A}.18)
$$
$$
R(2,7/2,\Gamma)=-{4\over 5\Gamma}{\partial \over
\partial \Gamma}R(2,5/2,\Gamma),
\eqno ({\rm A}.19)
$$
$$
R(3,7/2,\Gamma)=R(1,5/2,\Gamma)
-R(2,7/2,\Gamma)-{1\over 4}(1+\Gamma^{2})
R(1,7/2,\Gamma).
\eqno ({\rm A}.20)
$$
\vskip 0.3cm
\leftline {\bf Acknowledgments}
\vskip 0.3cm
\noindent
The author is much indebted to Ivan Avramidi and Alexander
Kamenshchik for scientific collaboration on heat-kernel
methods in quantum field theory. Correspondence with Andrei
Barvinsky, Stuart Dowker, Klaus Kirsten, 
Hugh Osborn and Dmitri Vassilevich was also very helpful.
\vskip 0.3cm
\leftline {\bf References}
\vskip 0.3cm
\item {[1]}
Branson T. P. and Gilkey P. B. (1990) {\it Commun. Part. Diff. Eqs.}
{\bf 15}, 245.
\item {[2]}
Gilkey P. B. (1995) {\it Invariance Theory, the Heat Equation
and the Atiyah-Singer Index Theorem} (Boca Raton: Chemical
Rubber Company).
\item {[3]}
Esposito G. (1998) {\it Dirac Operators and Spectral Geometry}
(Cambridge: Cambridge University Press).
\item {[4]}
Grubb G. (1996) {\it Func\-tio\-nal Cal\-cu\-lus 
o\-f Pseu\-do Dif\-fe\-ren\-ti\-al
Bo\-un\-da\-ry Pro\-ble\-ms} (Bos\-ton: Birk\-h\"{a}u\-ser).
\item {[5]}
G\"{u}nther P. and Schimming R. (1977) {\it J. Diff. Geom.}
{\bf 12}, 599.
\item {[6]}
Barvinsky A. O. (1987) {\it Phys. Lett.} {\bf B 195}, 344.
\item {[7]}
Moss I. G. and Poletti S. (1990) {\it Nucl. Phys.} 
{\bf B 341}, 155.
\item {[8]}
Esposito G. (1994) {\it Quantum Gravity, Quantum Cosmology 
and Lorentzian Geometries} (Berlin: Springer-Verlag).
\item {[9]}
Esposito G. and Kamenshchik A. Yu. (1995) 
{\it Class. Quantum Grav.} {\bf 12}, 2715.
\item {[10]}
Marachevsky V. N. and Vassilevich D. V. (1996) 
{\it Class. Quantum Grav.} {\bf 13}, 645.
\item {[11]}
Avramidi I. G., Esposito G. and Kamenshchik A. Yu. (1996)
{\it Class. Quantum Grav.} {\bf 13}, 2361.
\item {[12]}
Esposito G., Kamenshchik A. Yu. and Pollifrone G. (1997)
{\it Euclidean Quantum Gravity on Manifolds with Boundary}
(Dordrecht: Kluwer).
\item {[13]}
Esposito G. (1998) Non-Local Boundary Conditions in Euclidean
Quantum Gravity (GR-QC 9806057).
\item {[14]}
Luckock H. C. (1991) {\it J. Math. Phys.} {\bf 32}, 1755.
\item {[15]}
McAvity D. M. and Osborn H. (1991) {\it Class. Quantum Grav.}
{\bf 8}, 1445.
\item {[16]}
Moss I. G. and Silva P. J. (1997) {\it Phys. Rev.}
{\bf D 55}, 1072.
\item {[17]}
Avramidi I. G. and Esposito G. (1997) Gauge Theories on
Manifolds with Boundary (HEP-TH 9710048). 
\item {[18]}
Dowker J. S. and Kirsten K. (1997) {\it Class. Quantum Grav.}
{\bf 14}, L169.
\item {[19]}
Kirsten K. (1998) {\it Class. Quantum Grav.} {\bf 15}, L5.
\item {[20]}
Avramidi I. G. and Esposito G. (1998) {\it Class. Quantum Grav.}
{\bf 15}, 281.
\item {[21]}
Dowker J. S. and Kirsten K. (1998) The $a_{3/2}$ Heat-Kernel
Coefficient for Oblique Boundary Conditions (HEP-TH 9806168).
\item {[22]}
Avramidi I. G. and Esposito G. (1998) {\it Class. Quantum Grav.}
{\bf 15}, 1141.
\item {[23]}
Branson T. P. (1996) {\it Commun. Math. Phys.} {\bf 178}, 301.
\item {[24]}
Erdmenger J. (1997) {\it Class. Quantum Grav.} {\bf 14}, 2061.
\item {[25]}
Erdmenger J. and Osborn H. (1998) {\it Class. Quantum Grav.}
{\bf 15}, 273.
\item {[26]}
Avramidi I. G. (1997) {\it Phys. Lett.} {\bf B 403}, 280.
\item {[27]}
Avramidi I. G. (1998) {\it J. Math. Phys.} {\bf 39}, 2889.
\item {[28]}
Esposito G. (1997) {\it Phys. Rev.} {\bf D 56}, 2442.
\item {[29]}
Esposito G. and Kamenshchik A. Yu. (1998) Fourth-Order Operators
on Manifolds with Boundary (HEP-TH 9803036).
\item {[30]}
Eastwood M. and Singer I. M. (1985) {\it Phys. Lett.} 
{\bf A 107}, 73.
\item {[31]}
Schleich K. (1985) {\it Phys. Rev.} {\bf D 32}, 1889.
\item {[32]}
Lifshitz E. M. and Khalatnikov I. M. (1963) {\it Adv. Phys.}
{\bf 12}, 185.
\item {[33]}
DeWitt B. S. (1965) {\it Dynamical Theory of Groups and Fields}
(New York: Gordon and Breach).
\item {[34]}
Atiyah M. F., Patodi V. K. and Singer I. M. (1975) {\it Math.
Proc. Camb. Phil. Soc.} {\bf 77}, 43.
\item {[35]}
D'Eath P. D. and Halliwell J. J. (1987) {\it Phys. Rev.}
{\bf D 35}, 1100.
\item {[36]}
Schr\"{o}der M. (1989) {\it Rep. Math. Phys.} {\bf 27}, 259.
\item {[37]}
DeWitt B. S. (1967) {\it Phys. Rev.} {\bf 160}, 1113.
\item {[38]}
Avramidi I. G. and Branson T. P. (1998) Leading Symbols and
Representation Theory (work in preparation).
\item {[39]}
Avramidi I. G. and Branson T. P. (1998) Heat-Kernel Asymptotics
of Non-Diagonal Operators (work in preparation).
\item {[40]}
Avramidi I. G. (1991) {\it Int. J. Mod. Phys.} {\bf A 6}, 1693.
\item {[41]}
Misner C. W. (1957) {\it Rev. Mod. Phys.} {\bf 29}, 497.
\item {[42]}
Hawking S. W. (1979) in {\it General Relativity, an Einstein
Centenary Survey}, eds. S. W. Hawking and W. Israel
(Cambridge: Cambridge University Press) 746.
\item {[43]}
DeWitt B. S. (1984) in {\it Relativity, Groups and Topology II},
eds. B. S. DeWitt and R. Stora (Amsterdam: North Holland) 381.
\item {[44]}
Gibbons G. W. and Hawking S. W. (1993) {\it Euclidean Quantum Gravity}
(Singapore: World Scientific).

\bye